\documentclass[%
 reprint,
 superscriptaddress,
nofootinbib,
 amsmath,amssymb,
 aps,
]{revtex4-2}

\usepackage{graphicx}
\usepackage{dcolumn}
\usepackage{bbm}
\usepackage{hyperref}
\usepackage{xcolor}
\usepackage{braket}
\usepackage[normalem]{ulem}
\usepackage[T1]{fontenc}

\renewcommand{\eqref}[1]{Eq.~(\ref{#1})}

\renewcommand{\i}{\mathrm{i}}

\newcommand{\hc}{\mathrm{h.c.}}
\newcommand{\tr}{\mathrm{Tr}}

\begin{document}

\title{Bridging steady-state and time-domain descriptions of molecular electron transport}

\author{Thibaut Lacroix}
\affiliation{
Institut für Theoretische Physik \& IQST, Albert-Einstein-Allee 11, Universität Ulm, D-89081 Ulm, Germany
}

\author{Namgee Cho}
\affiliation{
Institut für Theoretische Physik \& IQST, Albert-Einstein-Allee 11, Universität Ulm, D-89081 Ulm, Germany
}

\author{Clemens Vittmann}
\affiliation{
Institut für Theoretische Physik \& IQST, Albert-Einstein-Allee 11, Universität Ulm, D-89081 Ulm, Germany
}

\author{James Lim}
\affiliation{
Institut für Theoretische Physik \& IQST, Albert-Einstein-Allee 11, Universität Ulm, D-89081 Ulm, Germany
}

\author{Susana F. Huelga}
\affiliation{
Institut für Theoretische Physik \& IQST, Albert-Einstein-Allee 11, Universität Ulm, D-89081 Ulm, Germany
}

\author{Martin B. Plenio}
\email{martin.plenio@uni-ulm.de}
\affiliation{
Institut für Theoretische Physik \& IQST, Albert-Einstein-Allee 11, Universität Ulm, D-89081 Ulm, Germany
}

\begin{abstract}
Electron transmission from an input electrode, through a molecular system, to an output electrode has been widely studied using the steady-state non-equilibrium Green's function (NEGF) method. Recently, the wave packet method, which provides access to the transient dynamics of electrons as well as internal molecular degrees of freedom, has been employed to investigate enantiospecific electron transport through chiral molecules. 
In this work, we derive the quantitative relation between the transmission of a finite-size wave packet and the energy-resolved NEGF transmission, showing that the former corresponds to a spectral average of the latter weighted by the wave packet energy distribution.
Exploiting this correspondence, we construct non-Gaussian auxiliary wave packets whose spectral weight encodes the Landauer energy-window, allowing current-voltage characteristics to be obtained directly from time propagation.
We further show that the correspondence extends to spin-resolved transport in a spin-phonon model of chirality-induced spin selectivity.
\end{abstract}

\maketitle

\section{Introduction}\label{sec:introduction}

Electron transport through molecular or nanoscale systems has been extensively investigated in a variety of experiments, including chirality-induced spin selectivity (CISS)~\cite{xie_spin_2011, kettner_spin_2015, aragones_measuring_2017, mishra_spin-dependent_2019, mondal_spin_2021}, molecular rectification~\cite{guo2016molecular, chen2017molecular}, single-molecule conductance~\cite{xu_measurement_2003, chen_measurement_2007}, thermoelectric effects in molecular junctions~\cite{reddy_thermoelectricity_2007}, and quantum interference phenomena in conjugated systems~\cite{guedon_observation_2012}. The non-equilibrium Green's function method has been widely employed to analyze and interpret these phenomena~\cite{taylor2001ab, Brandbyge2002Density, paulsson_thermoelectric_2003, Stadler_2008, hartle_quantum_2011, guo_spin-selective_2012, guo_spin-dependent_2014, gutierrez_spin-selective_2012, gersten_induced_2013}. Here, a molecular system is typically placed between two electrodes, and the electron transmission from the input electrode, through the molecule, to the output electrode is examined. In the NEGF formalism, the Hamiltonian of the electrodes is not treated explicitly. Instead, their influence on the electron transmission is incorporated via the corresponding Green's functions and self energies~\cite{datta_quantum_2005, ryndyk_green_2009, cuevas_molecular_2010, camsari_nonequilibrium_2023}. This avoids the need to treat macroscopic electrodes directly, thereby limiting the computational cost to the dimension of the molecular system's Hamiltonian. 
However, this steady-state NEGF approach does not provide access to the transient dynamics of electrons and other physical degrees of freedom coupled to the electron motion, such as phonon modes, which renders the interpretation of the simulated results inherently challenging.

Recently, the wave packet (WP) method has been employed to investigate the mechanisms of enantiospecific electron dynamics in chiral molecular systems, with electrodes explicitly included in the simulations, along with a finite-sized electron wave packet~\cite{vittmann_interface-induced_2022,vittmann_spin-dependent_2023,cho_chirality-induced_2026}. Although the total dimension of the Hamiltonian treated explicitly in WP is larger than in NEGF, WP deals with pure-state dynamics of an electron wave packet and does not require matrix inversion, such as that involved in the evaluation of Green's functions in NEGF. In addition, absorptive layers can be introduced in the electrodes, allowing the Hilbert space of the WP model to be truncated with controllable boundary error. 
Importantly, the WP approach provides access to the transient dynamics of electrons and other internal degrees of freedom within the molecule, enabling detailed identification of the mechanisms underlying the simulated results.
However, it has not yet been demonstrated what the explicit quantitative correspondence is between stationary NEGF observables and finite-size wave packet propagation, and whether this correspondence can be exploited to calculate experimentally relevant, energy-integrated, transport quantities directly in the time-domain.
In this work, we establish an operational correspondence between the two frameworks, and consider a molecular tight-binding model and a spin-phonon model for CISS~\cite{vittmann_spin-dependent_2023} as examples to demonstrate that the steady-state NEGF and WP methods yield quantitatively consistent
asymptotic quantities, such as electron transmittance, current-voltage characteristics, and spin polarization.
We thus establish a close connection between studies employing NEGF~\cite{fransson_chirality-induced_2019,du_vibration-enhanced_2020,fransson_vibrational_2020} and those employing the WP method~\cite{vittmann_interface-induced_2022,vittmann_spin-dependent_2023,cho_chirality-induced_2026}.
This is particularly important in the context of CISS, where differences between theoretical predictions are sometimes attributed to different physical mechanisms, although they may also reflect differences in model assumptions or in the representation of transport observables.

This paper is organized as follows. Section~\ref{sec:NEGF_TT} introduces a two-terminal setup to explain the basic concepts of the NEGF method. Section~\ref{sec:NEGF_MT} discusses a multi-terminal setup, showing how the NEGF method can be applied to study enantiospecific electron dynamics in the presence of spin-phonon coupling. Section~\ref{sec:WP} presents the WP method, including strategies to improve computational efficiency and stability via absorptive layers. 
Section~\ref{sec:results} establishes the asymptotic equivalence of the two frameworks, verifies it numerically, and exploits it to compute non-trivial energy-integrated transport observables, such as current and spin-polarization, in the time-domain.
Finally, Sec.~\ref{sec:conclusions} provides the conclusions.

\section{Non-equilibrium Green's function method}\label{sec:NEGF}

The non-equilibrium Green's function formalism can be derived in different ways, such as from many-body perturbation theory or from the Schr{\"o}dinger equation with open boundary conditions~\cite{keldysh2024diagram, kadanoff_quantum_2018, datta_non-equilibrium_2015}. Here, we discuss how the Green's function method can be applied to the scattering problem of a single-electron system involving several physical degrees of freedom, such as the electron's spin and electron-phonon coupling, as considered in Ref.~\cite{vittmann_spin-dependent_2023}.

\subsection{Two-Terminal Problem}\label{sec:NEGF_TT}

\begin{figure}
    \centering
    \includegraphics[width=\columnwidth]{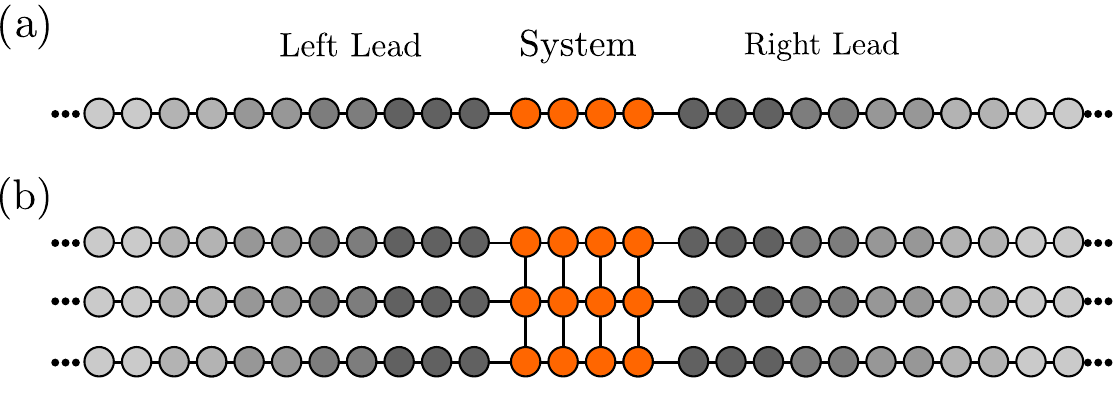}
    \caption{(a) Schematic of a two-terminal setup where a molecular system is coupled to an input (left) and an output (right) lead. There is no direct coupling between the two leads. (b) Schematic of a multi-terminal setup where the molecular system is coupled to multiple non-interacting input and output leads.}
    \label{fig:leads}
\end{figure}

We start with a two-terminal setup in which an electron is transferred from a left lead, through a molecular system, to a right lead, as schematically illustrated in Fig.~\ref{fig:leads}(a). To introduce the basic concepts of the NEGF method, we only consider the electron's position degrees of freedom, and neglect other degrees of freedom, such as the electron’s spin and phonon modes coupled to the electron, which will be addressed in Sec.~\ref{sec:NEGF_MT}.

In the NEGF formalism, the leads are often modeled as semi-infinite chains with uniform site energies $\epsilon_j$ and inter-site couplings $t_j$ for the left ($j=L$) and right ($j=R$) leads, described by the lead Hamiltonians~\cite{datta_quantum_2005, ryndyk_green_2009, camsari_nonequilibrium_2023}
\begin{align}
    H_{j\in \{L,R\}} &= \sum_{n=1}^{\infty} \epsilon_j a_{j,n}^\dagger a_{j,n} + t_j(a_{j,n}^\dagger a_{j,n+1} + \hc) ,\label{eq:H-leads}
\end{align}
where $a_{j,n}^\dagger$ and $a_{j,n}$ denote the creation and annihilation operators, respectively, for an electron at site $n$ of lead $j$.
Throughout this work, we set $\hbar=1$. 
The eigenvalues of the leads are given by $E_j(k) = \epsilon_j + 2t_j\cos(k)$ with $k \in [0,\ \pi]$ and the corresponding eigenstates are $\psi_k \propto \sin(k n)$. The Hamiltonian of a molecular system consisting of $N$ sites may be defined similarly as
\begin{align}
    H_{S} &= \sum_{n=1}^{N} \epsilon_n c_{n}^\dagger c_{n} + \sum_{n=1}^{N-1} t_{n}(c_{n}^\dagger c_{n+1} + \hc),\label{eq:H_S_no_spin_phonon}
\end{align}
with site energies $\epsilon_n$ and inter-site couplings $t_{n}$, where $c_{n}^\dagger$ and $c_{n}$ represent the creation and annihilation operators, respectively, for an electron at site $n$ within the molecule. We assume that the first site of the left (right) lead is coupled to the first (last) site of the molecular system via the interfacial coupling $g_L$ ($g_R$), described by
\begin{align}
    H_{jS} = g_{j} a_{j,1}^\dagger c_j,
\end{align}
with $j\in\{L,R\}$, where $c_L = c_1$ and $c_R = c_N$.

Within one-electron theory, the scattering problem of the two-terminal setup is formulated as the eigenvalue equation of the total Hamiltonian
\begin{align}\left(
    \begin{pmatrix}
        H_L& H_{LS}& 0\\
        H_{LS}^\dagger & H_S& H_{RS}^\dagger\\
        0& H_{RS}& H_R
    \end{pmatrix}
    -E\right)
    \begin{pmatrix}
        \psi_L^{(\text{in})}  + \psi_L^{(r)}\\
        \psi_S\\
        \psi_R^{(t)}
    \end{pmatrix} 
    & = 0, 
\end{align}
where $\psi_L^{(\text{in})}\propto \sin(k n)$ is the input electron wave described by the eigenstate of the left lead with energy $E=\epsilon_L + 2t_L\cos(k)$, $\psi_L^{(r)}$ denotes the reflected wave from the molecular system, and $\psi_R^{(t)}$ the transmitted wave through the molecule. The eigenvalue equation can be rewritten as~\cite{ryndyk_green_2009}
\begin{align}
    \psi_L^{(r)} &= G_L(E) H_{LS} G_S(E) H_{LS}^\dagger \psi_L^{(\text{in})}, \label{eq:SP1}\\
    \psi_S &= G_S(E) H_{LS}^\dagger\psi_L^{(\text{in})}, \label{eq:SP2}\\
    \psi_R^{(t)} &= G_R(E) H_{RS}G_S(E) H_{LS}^\dagger\psi_L^{(\text{in})}, \label{eq:SP3}
\end{align}
where $G_L(E)$ and $G_R(E)$ denote the retarded Green's functions of the leads
\begin{align}
    G_{j\in\{L,R\}}(E) = \lim_{\varepsilon\rightarrow 0^+} (E - H_j + \i \varepsilon)^{-1}\ ,
\end{align}
and $G_S(E)$ is the retarded Green's function of the molecular system
\begin{align}
    G_S(E) &= \left(E - H_S - \Sigma_L(E) - \Sigma_R(E)\right)^{-1}, \label{eq:Gs}
\end{align}
where $\Sigma_L(E)$ and $\Sigma_R(E)$ are the self-energies induced by the interaction between the leads and the molecular system
\begin{align}
    \Sigma_{j\in\{L,R\}}(E) &= H_{jS}^{\dagger}G_j(E)H_{jS}.
\end{align}
For the semi-infinite chains defined in \eqref{eq:H-leads}, the self-energies can be expressed analytically as
\begin{align}
    \Sigma_j(E) = \frac{|g_j|^2}{|t_j|}&\Big(\kappa_j - \sqrt{\kappa_j^2-1}\left(\Theta(\kappa_j-1)-\Theta(-\kappa_j-1)\right)\nonumber\\
    &\quad-{\i}\sqrt{1-\kappa_j^2}\Theta(1-|\kappa_j|)\Big)c_j^\dagger c_j\ ,
\end{align}
where $\kappa_j = (E-\epsilon_j)/(2|t_j|)$, and $\Theta$ is the Heaviside step-function with $\Theta(x)=1$ for $x\ge 0$ and $\Theta(x)=0$ otherwise.

From the formal solution of the scattering problem in Eqs.~(\ref{eq:SP1})-(\ref{eq:SP3}), written in terms of the Green's functions, the transmission probability from the left lead, through the molecular system, to the right lead can be expressed as~\cite{ryndyk_green_2009}
\begin{align}
    T_{L\rightarrow R}(E) = \tr\left[\Gamma_L(E) G_S^\dagger(E) \Gamma_R(E) G_S(E) \right],
\end{align}
where $\Gamma_{L}(E)$ and $\Gamma_{R}(E)$ denote the level-broadening operators, defined as
\begin{align}
    \Gamma_{j\in\{L,R\}}(E) = \i (\Sigma_j(E) - \Sigma_j^\dagger(E)).
\end{align}

\subsection{Multi-Terminal Problem}\label{sec:NEGF_MT}

So far, we have considered the two-terminal problem, where a molecular system interacts with two independent leads. The NEGF formalism can be generalized to the multi-terminal case, in which the molecule is coupled to multiple non-interacting leads. For instance, in the case of two left and two right leads, the Hamiltonian takes the form
\begin{align}
    H &=
    \begin{pmatrix}
            H_{L_1}& 0 & H_{L_1 S} & 0 & 0 \\
            0 & H_{L_2} & H_{L_2 S} & 0 & 0 \\
            H_{L_1 S}^\dagger & H_{L_2 S}^\dagger & H_{S} & H_{R_1 S}^\dagger & H_{R_2 S}^\dagger \\
            0 & 0 & H_{R_1 S} & H_{R_1} & 0 \\
            0 & 0 & H_{R_2 S} & 0 & H_{R_2}
    \end{pmatrix},
\end{align}
where $H_{L_1}$, $H_{L_2}$, $H_{R_1}$, $H_{R_2}$ denote the Hamiltonians of independent semi-infinite chains without inter-chain couplings, each of which is coupled to the molecular system, as schematically illustrated in Fig.~\ref{fig:leads}(b). In this case, the retarded Green's function for the molecular system is expressed as
\begin{align}
    G_S &= \left(E - H_S - \sum_k (\Sigma_{L_k} + \Sigma_{R_k})\right)^{-1}, \label{eq:Gs_general}
\end{align}
which includes the self-energies arising from all the leads, whose dependence on the energy $E$ is not written explicitly for simplicity,
\begin{align}
    \Sigma_{\xi} &= \frac{|g_\xi|^2}{|t_\xi|}\left(\kappa_\xi - \sqrt{\kappa_\xi^2-1}\left(\Theta(\kappa_\xi-1)-\Theta(-\kappa_\xi-1)\right)\right.\nonumber\\
    &\qquad\qquad\,\,\left.- {\i}\sqrt{1-\kappa_\xi^2}\Theta(1-|\kappa_\xi|)\right)c_\xi^\dagger c_\xi\ ,
\end{align}
where $\xi\in\{L_k, R_k | k=1,2,\cdots\}$, $\kappa_\xi = (E-\epsilon_\xi)/(2|t_\xi|)$, and $c_\xi^\dagger$ ($c_\xi$) is the creation (annihilation) operator for an electron at the molecular site coupled to the lead $\xi$. Note that the site energies $\epsilon_\xi$ and the inter-site couplings $t_\xi$ of the leads are generally independent. The transmission probability from a given left lead $L_k$ to a particular right lead $R_l$ is
\begin{align}
    T_{L_k\rightarrow R_l}(E) = \tr\left[\Gamma_{L_k} G_S^\dagger \Gamma_{R_l} G_S \right],
\end{align}
where the level-broadening operators are given by
\begin{align}
    \Gamma_{\xi} = \i (\Sigma_\xi - \Sigma_\xi^\dagger).\label{eq:Gamma_general}
\end{align}

The NEGF formalism for the multi-channel problem can be employed to simulate electron transport from the left to the right lead in the presence of additional physical degrees of freedom, such as the electron’s spin, and phonon modes that induce fluctuations in the electronic parameters, as considered in Ref.~\cite{vittmann_spin-dependent_2023}. When spin-orbit and spin-phonon couplings are absent in the leads but present only within the molecular system, the Hamiltonians of the left ($j=L$) and right lead ($j=R$) are given by
\begin{align}
    H_{j} &= \sum_{n=1}^{\infty}\sum_{s\in\{\uparrow,\downarrow\}} \epsilon_j a_{j,n,s}^\dagger a_{j,n,s} + t_j(a_{j,n,s}^\dagger a_{j,n+1,s} + \hc)\nonumber\\
    &\quad+\omega b^\dagger b,
\end{align}
where $a_{j,n,s}^\dagger$ ($a_{j,n,s}$) creates (annihilates) an electron with spin $s\in\{\uparrow,\downarrow\}$ (e.g., a spin-up or spin-down eigenstate of the Pauli operator $\sigma_z$) at site $n$ of lead $j\in\{L,R\}$. The last term describes the vibrational energy of a harmonic molecular phonon mode with frequency $\omega$, where $b^\dagger$ ($b$) is the phonon creation (annihilation) operator.
Thus, when the electron is in lead $j\in\{L, R\}$ the states $\{\ket{m}\}_{m\in\mathbb{N}}$ label its asymptotic scattering channels.
We decompose the Hamiltonian of lead $j$ into multiple semi-infinite chains depending on spin and phonon states $\ket{s,m}$, defined by $b^\dagger b\ket{m}=m\ket{m}$ with non-negative integer $m$, leading to the Hamiltonians of the individual leads $j_{sm}$ without inter-lead couplings,
\begin{widetext}
\begin{align}
    H_{j_{sm}} = \left( m\omega+\sum_{n=1}^{\infty}\left(\epsilon_{j} a_{j,n,s}^\dagger a_{j,n,s}+ t_j(a_{j,n,s}^\dagger a_{j,n+1,s} + \hc)\right) \right)\otimes \ket{m}\bra{m},
\end{align}
\end{widetext}
satisfying $H_j = \sum_{s\in\{\uparrow,\downarrow\}}\sum_{m=0}^{\infty} H_{j_{sm}}$. 
When the interfacial coupling between the leads and the molecule does not depend on spin and phonon states, the interaction Hamiltonian reads
\begin{align}
    H_{j_{sm}S} = g_{j}a_{j,1,s}^{\dagger}c_{j,s} \otimes \ket{m}\bra{m},
\end{align}
where $c_{L,s}=c_{1,s}$ and $c_{R,s}=c_{N,s}$. This multi-channel NEGF formalism enables the computation of the transmission probability from the left lead to the right lead as a function of the total energy $E$ of the electron-phonon system, conditioned on the spin-phonon states via Eqs.~(\ref{eq:Gs_general})-(\ref{eq:Gamma_general}), namely
\begin{align}
    T_{L_{sm}\rightarrow R_{s'm'}}(E) = \tr\left[\Gamma_{L_{sm}} G_S^\dagger \Gamma_{R_{s'm'}} G_S \right],\label{eq:transmission_function}
\end{align}
where an electron with initial spin $s$ is transmitted to the right lead with final spin $s'$, while the phonon state changes from $\ket{m}$ to $\ket{m'}$, due to the spin-orbit and/or spin-phonon couplings in the molecular Hamiltonian $H_S$. We note that the multi-terminal NEGF approach can also be applied to other physical models, such as those involving the electron’s orbital angular momentum~\cite{cho_chirality-induced_2026}, instead of the phonon mode.

In this work, we consider the spin-phonon coupling from Ref.~\cite{vittmann_spin-dependent_2023} as an example 
\begin{align}
    H_{S} &= \sum_{n=1}^{N} \epsilon_n c_{n}^\dagger c_{n} + \sum_{n=1}^{N-1} t_{n}(c_{n}^\dagger c_{n+1} + \hc)+\omega b^\dagger b \nonumber\\
    &\quad+{\rm i}\frac{\alpha}{2}\left(b + b^\dagger\right)\sum_{n=1}^{N-1} (c_{n}^\dagger (\sigma_n + \sigma_{n+1}) c_{n+1} + \hc) , \label{eq:H_S_spin_phonon}
\end{align}
with $\alpha$ quantifying the spin-phonon coupling strength, where $c_n^\dagger = (c_{n,\uparrow}^\dagger, c_{n,\downarrow}^\dagger)$ denotes the spinor row vector composed of the creation operators for an electron in spin-up and spin-down states (in the Pauli $\sigma_z$ basis) at molecular site $n$. Here, $\sigma_n$ is defined as
\begin{align}
    \sigma_n=\frac{\sin(\beta s_n )}{D}
    \begin{pmatrix}
        R & {\rm i}P{\rm e}^{-{\rm i}s_n/D} \\ 
        -{\rm i}P{\rm e}^{{\rm i}s_n/D} & -R
    \end{pmatrix},
\end{align}
with $s_n=2\pi D (n-1)/N$, where $R$ and $2\pi P$ denote the radius and pitch of a helical molecule, respectively, with $D=\sqrt{R^2+P^2}$, and $2\pi/\beta$ represents the wavelength of a delocalized phonon mode of the molecule. 
This model considers only a single vibrational mode but can be easily extended to include multiple vibrational modes.
Here, we assume that the electron is initially in a random mixture of spin-up and spin-down states, and that the phonon mode is in a thermal state. 
The final spin $s'$-dependent transmission probability $T_{s'}$ can then be computed by averaging the transmission function in Eq.~(\ref{eq:transmission_function}) over all possible values of the initial spin $s$ and phonon quantum numbers $m$ and $m'$, weighted by the Boltzmann factor $\propto \exp(-m\omega/k_B T)$ at temperature $T$. 
The spin polarization is defined as ${\rm SP}=(T_\uparrow-T_\downarrow)/(T_\uparrow+T_\downarrow)$.

\section{Wave packet method}\label{sec:WP}

\begin{figure}
    \centering
    \includegraphics[width=\columnwidth]{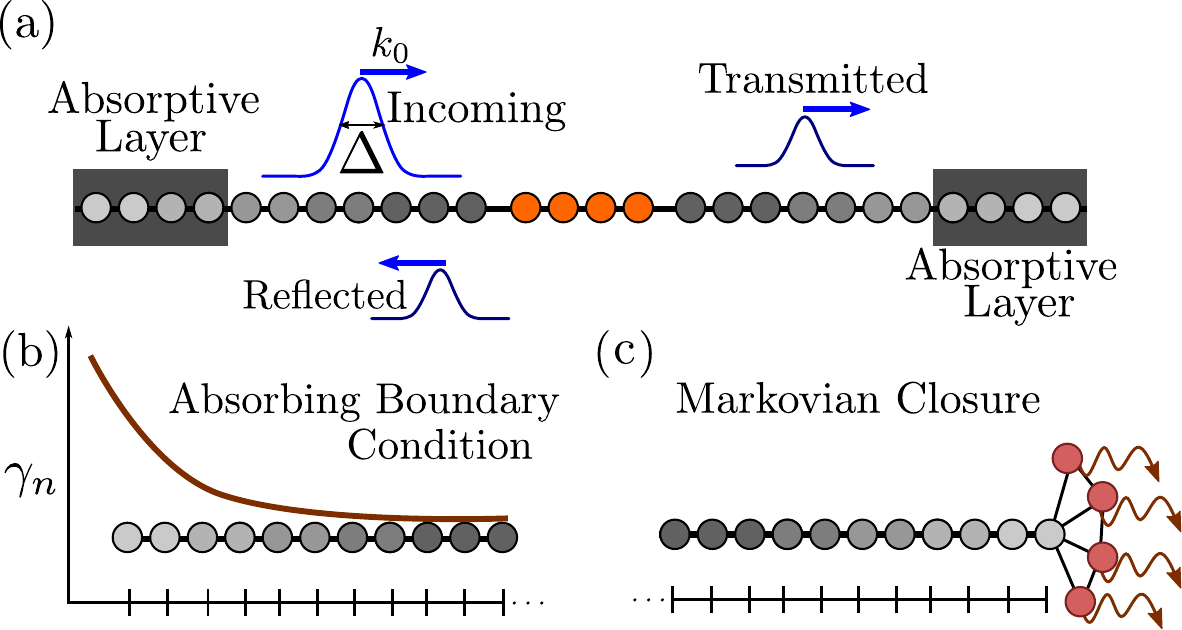}
    \caption{(a) Schematic of a molecular system (orange circles) coupled to an input (left) and an output (right) lead. The initial state, with wavevector $k_0$ and width $\Delta$, is placed in the input lead and scattered by the molecular system, resulting in reflected and transmitted waves. Each lead consists of a finite number of sites, with the end site coupled to an absorptive layer that prevents the reflected or transmitted wave from re-entering the molecular system. The absorptive layer can be modeled as (b) a linear chain with gradually increasing Lindblad damping rates $\gamma_n$, or (c) a small number of sites (e.g., 6-10) with a carefully tuned set of damping rates, optimized to efficiently absorb the reflected or transmitted wave, known as the Markovian closure~\cite{nuseler_fingerprint_2022,vittmann_spin-dependent_2023}.}
    \label{fig:ABC-MC}
\end{figure}

In contrast to the NEGF formalism, in the wave packet method, the Hamiltonians of the leads are explicitly included in the simulations. The input electron state is represented as a finite-sized wave packet in the site basis $\{\ket{n}\}$, for example a Gaussian wave packet,
\begin{align}
    \ket{\psi(0)}=\sum_{n} A {\rm e}^{\i k_0 (n-n_0)-(n-n_0)^2/(2\Delta^2)}\ket{n}\otimes \ket{\chi}, \label{eq:Gaussian_initial_state}
\end{align}
where $A$ is the normalization factor, and $\ket{\chi}$ denotes the quantum state of physical degrees of freedom other than position (i.e., sites), such as the electron's spin and phonon states, e.g., $\ket{\chi}=\ket{s,m}$ with $s\in\{\uparrow,\downarrow\}$ and $m$ denoting the number of phonons. For a given width $\Delta$, the center $n_0$ is chosen such that the wave packet is well localized inside the left lead, with negligible amplitude in the molecular region, as schematically illustrated in Fig.~\ref{fig:ABC-MC}(a). The average energy of this initial state is determined by the wavevector $k_0$. For a homogeneous semi-infinite chain representing the left lead, the average energy of the electron follows from the dispersion relation $E(k_0)=\epsilon_L+2t_L\cos(k_0)$. The energy uncertainty of the initial state decreases as the width $\Delta$ increases. We note that the energy $E(k_0)$ is independent of the sign of $k_0$. In simulations, we choose the sign such that the initial wave packet propagates toward the molecular system.

The propagation of the electron wave packet can be dynamically simulated by solving the Schr{\" o}dinger equation, for example using the fourth-order Runge-Kutta method. 
This approach allows one to monitor the time- and position-dependent dynamics of the electron, as well as other physical degrees of freedom in the model, such as the electron's spin or phonon modes. 
Access to such transient dynamics is valuable for identifying the mechanisms underlying electron transmittance through the molecular system~\cite{vittmann_interface-induced_2022,vittmann_spin-dependent_2023,cho_chirality-induced_2026}. 
In particular, it enables one to determine whether important physical effects occur at the interface between the leads and the molecule~\cite{vittmann_interface-induced_2022} or within the molecular region itself~\cite{vittmann_spin-dependent_2023,cho_chirality-induced_2026}. 
In contrast, the steady-state NEGF method summarized in Sec.~\ref{sec:NEGF} does not directly resolve such transient dynamics.
Propagation of a physical wave packet therefore provides complementary time-domain information on when and where the electronic, spin, and phonon degrees of freedom become dynamically correlated.

The transmittance of the electron is computed by simulating the wave packet dynamics until the reflected and transmitted components, generated through scattering by the molecular system, are well separated from the molecular region, leaving negligible population within the molecule. The transmission probability is then obtained by evaluating the total population in the right lead. In simulations, both the left and right leads are truncated to a finite number of sites. The leads must be sufficiently large to prevent the reflected or transmitted waves from reaching the truncated boundaries, being reflected, and re-entering the molecular region. This requirement sets a limit on the simulation time for a given length of the leads: as the propagation time increases, the leads must be extended to avoid the finite-size effects.

Alternatively, absorptive layers can be included at the ends of the leads so that the reflected and transmitted waves decay before re-entering the molecular region. In this case, the simulation only needs to run long enough for the reflected and transmitted components to decay to negligible amplitudes. The absorptive layers can be implemented in several ways: (i) As a linear chain coupled to the truncated site of a lead, as schematically illustrated in Fig.~\ref{fig:ABC-MC}(b), where Lindblad damping rates $\gamma_n$ are assigned to each site and gradually increase along the absorptive layer. This gradual increase ensures that the time-evolution operator changes smoothly with position, preventing undesired reflections by the absorptive layer that could propagate back to the molecular region~\cite{riss_investigation_1996, muga_complex_2004}. (ii) As a small number of sites (e.g., 6-10) independently coupled to the truncated site of a lead, as illustrated in Fig.~\ref{fig:ABC-MC}(c). By using a carefully chosen set of finite Lindblad damping rates (which do not increase gradually), this so-called Markovian closure can efficiently absorb the electron wave reaching the truncated site of the lead~\cite{nuseler_fingerprint_2022,vittmann_spin-dependent_2023}. (iii) As an absorptive layer whose sites are evolved with a modified Schr{\" o}dinger equation that only allows for outgoing waves~\cite{antoine_review_2007, nagel_review_2009, wu_absorbing_2020}.

For the cases (i) and (ii), the Lindblad damping can be implemented by introducing a non-Hermitian term $K$ in the Schr{\" o}dinger equation
\begin{align}
    \frac{d}{dt}\ket{\psi(t)}&=-{\rm i}(H+K)\ket{\psi(t)},\\
    K&=-\frac{{\rm i}}{2}\sum_{n}\sum_{\chi'}\gamma_n \ket{n,\chi'}\bra{n,\chi'},
\end{align}
where $H$ denotes the total Hamiltonian of the leads and the molecule, and $\gamma_n$ is the damping rate at site $n$ of the absorptive layer. The non-Hermitian term $K$ reduces the norm of the quantum state $\ket{\psi(t)}$ when it has amplitude within the absorptive layer, eventually causing the state to vanish, i.e., $\langle \psi(t)|\psi(t)\rangle\rightarrow 0$ as $t\rightarrow \infty$. The transmission probability, conditioned on the final quantum state $\ket{\chi'}$ of the physical degrees of freedom other than position, such as the electron's spin and phonon states, is given by
\begin{align}
    T_{\chi\rightarrow \chi'}=\sum_{n}\gamma_n \int_{0}^\infty dt \,|\langle n,\chi'|\psi(t)\rangle|^2,
\end{align}
where $n$ labels the sites of the absorptive layer coupled to the right lead.

For the spin-phonon model in Eq.~(\ref{eq:H_S_spin_phonon}), we consider the Gaussian initial state in Eq.~(\ref{eq:Gaussian_initial_state}) with $\ket{\chi}=\ket{s,m}$, and compute the average of the transmittance $T_{(s,m)\rightarrow(s',m')}$ over all possible $(s,m,m')$, weighted by the Boltzmann factor $\propto {\rm exp}(-m\omega/k_B T)$, yielding the final spin $s'$-dependent transmittance $T_{s'}$, similar to the NEGF method. The spin polarization is also defined as ${\rm SP}=(T_\uparrow-T_\downarrow)/(T_\uparrow+T_\downarrow)$.

\section{Results}\label{sec:results}

In this section, we establish the connection between the steady-state, energy-resolved NEGF description and the time-domain wave-packet description. 
In particular, we show in Sec.~\ref{sec:results_trans} that the asymptotic WP propagation corresponds to an energy-weighted average of the stationary NEGF transmission over an energy distribution determined by the spatial profile of the initial WP.
This provides an operational connection between the two frameworks that we use in Sec.~\ref{sec:results_current} to construct an initial state whose asymptotic transmitted norm is used to compute a desired energy-integrated transport quantity, namely the Landauer current.
Finally, by computing the spin polarization of the spin-phonon model introduced in Sec.~\ref{sec:NEGF_MT}, we show in Sec.~\ref{sec:results_SP} that this equivalence still holds when the molecular region contains additional degrees of freedom.

\subsection{Transmittance}\label{sec:results_trans}

In the WP framework, the initial state supported in the input (left) lead can be expressed in $k$-space as
\begin{align}
    \ket{\psi(0)} = \int_{0}^{\pi} dk A_k \sum_{n}{\rm e}^{{\rm i}kn}\ket{n},
\end{align}
where the amplitudes $A_k$ determine the frequency spectrum of the WP.
The asymptotic transmission probability $T_{W}$ in the WP framework is then expressed in terms of the NEGF transmittance $T_{N}(k)$ as
\begin{align}
    T_{W} &= \int_{0}^{\pi} dk |A_k|^2 T_{N}(k)\ .\label{eq:WP_NEGF_k}
\end{align}
Using the dispersion relation $E=\epsilon_L+2 t_L\cos(k)$ of the input lead, \eqref{eq:WP_NEGF_k} can be rewritten in the energy domain as
\begin{align}
    T_{W} &= \int_{E_{\rm min}}^{E_{\rm max}}dE' \frac{|A_{k(E')}|^2}{2|t_L \sin(k(E'))|} T_{N}(E')\ ,\label{eq:WP_NEGF_E}
\end{align}
where $E_{\rm min}=\epsilon_L-2|t_L|$, $E_{\rm max}=\epsilon_L+2|t_L|$, and $k(E')=\cos^{-1}((E'-\epsilon_L)/(2t_L))$. 
This means that the asymptotic WP transmittance $T_W$ corresponds to an energy-weighted average of the stationary NEGF transmittance $T_N(E')$.
The weights are determined by the spectral content of the incoming WP which in turn is determined by the spatial profile of the WP.

Considering the initial Gaussian state in \eqref{eq:Gaussian_initial_state}, its spectrum is given by
\begin{align}
    A_k = \left(\frac{\Delta}{\sqrt{\pi}}\right)^{\frac{1}{2}} \mathrm{e}^{-\mathrm{i}kn_0 - \Delta^2(k-k_0)^2/2},
\end{align}
which is valid provided that $A_k$ is predominantly contained within $k\in[0,\pi]$ due to a sufficiently small width $1/\Delta$ in $k$-space.
Using \eqref{eq:WP_NEGF_E}, the transmittance $T_{W}(E)$, where $E=\epsilon_L+2t_L\cos(k_0)$, is expressed in terms of the NEGF transmittance as
\begin{align}
    T_{W}(E) &= \frac{\Delta}{\sqrt{\pi}}\int_{E_{\rm min}}^{E_{\rm max}}dE' \frac{{\rm e}^{-\Delta^2(k(E')-k_0)^2}T_{N}(E')}{2|t_L \sin(k(E'))|}\ , \label{eq:WP_NEGF_Gaussian}
\end{align}
where the prefactor $\Delta/\sqrt{\pi}$ ensures normalization, i.e. $T_{W}(k_0) = 1$ when $T_{N}(k) = 1$ for all $k$, provided that $|A_k|^2 \propto \mathrm{e}^{-\Delta^2(k-k_0)^2}$ is well contained within $k\in[0,\pi]$ due to a sufficiently small width $1/\Delta$ in $k$-space. 

\begin{figure}
    \centering
    \includegraphics[width=\columnwidth]{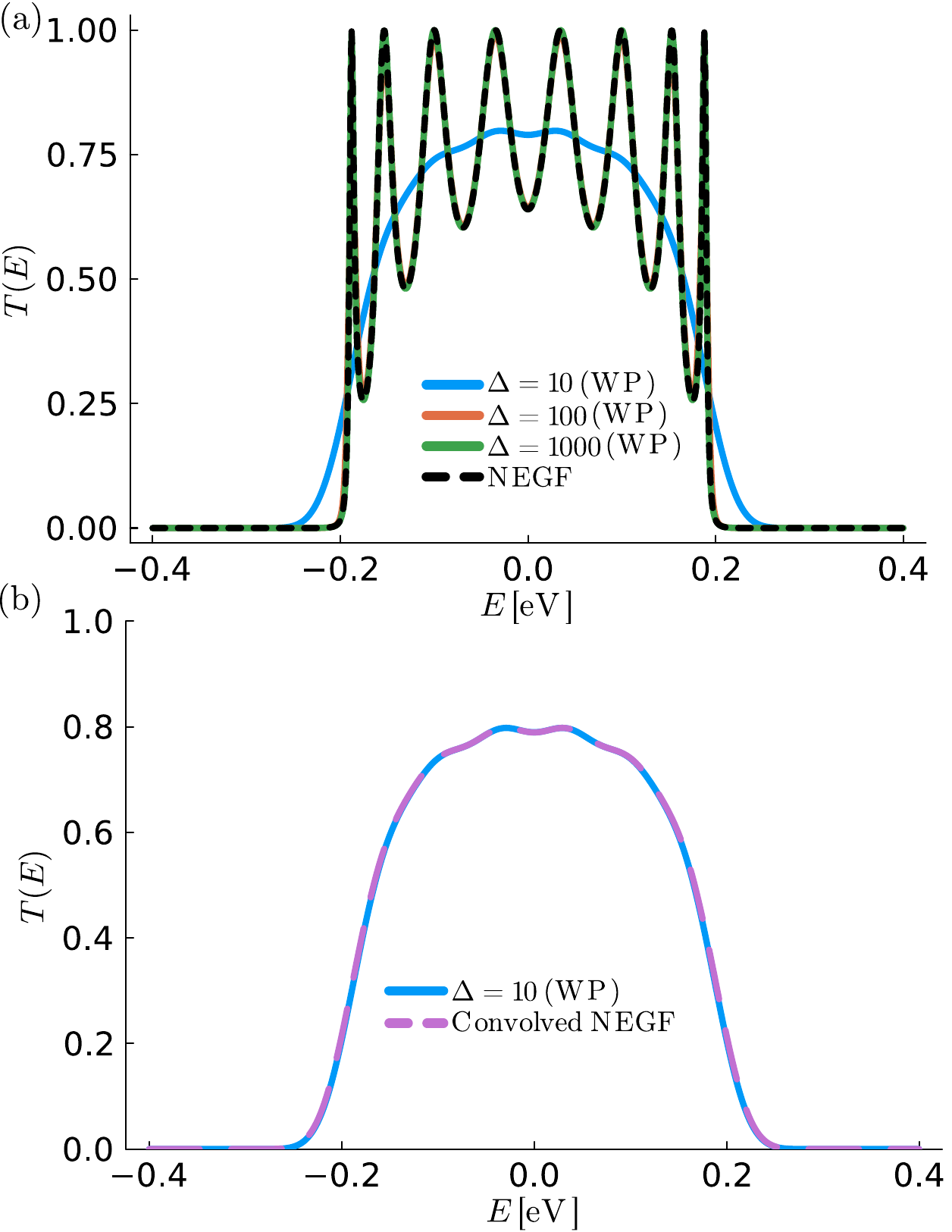}
    \caption{(a) From WP to NEGF: Transmittance of an electron in a two-terminal setup where the electron's spin and phonon mode are neglected. NEGF results are shown as a black dashed line, while WP results with controlled width $\Delta \in \{10, 100, 1000\}$ of the initial state are shown as colored solid lines. For $\Delta \gg 10$, the NEGF and WP results are well-matched. (b) From NEGF to WP: WP results with ${\Delta = 10}$ can be reproduced by averaging the NEGF results with a Gaussian weight (see the main text), shown as a purple dashed line. In the simulations, we consider $\epsilon_L = \epsilon_R = \epsilon_n = 0$, $t_L = t_R = 0.2\, \mathrm{eV}$, $t_n = 0.1\, \mathrm{eV}$, and $g_L = g_R = 0.2\, \mathrm{eV}$.}
    \label{fig:WP-electron}
\end{figure}

In Fig.~\ref{fig:WP-electron}, we consider the transmission of an electron through a linear molecular chain consisting of $N = 10$ sites, described by Eq.~(\ref{eq:H_S_no_spin_phonon}), in which both the electron's spin and the phonon mode are neglected. 
We assume that the leads and the molecule have identical site energies, $\epsilon_L = \epsilon_R = \epsilon_n = 0$, and consider the following inter-site couplings: $t_L = t_R = 0.2\ \mathrm{eV}$ and  $t_n = 0.1\,{\rm eV}$ for all molecular sites $n$. 
The interfacial couplings are set to $g_L = g_R = 0.2\ \mathrm{eV}$. 
This model can be simulated using the NEGF method for the two-terminal setup, as summarized in Sec.~\ref{sec:NEGF_TT}, which yields the transmittance as a function of the electron energy $E$, shown as a black dashed line in Fig.~\ref{fig:WP-electron}(a). 
The transmittance computed by the WP method with a controlled width $\Delta \in \{10,100,1000\}$ of the initial Gaussian state (see Eq.~(\ref{eq:Gaussian_initial_state})) are shown as colored solid lines. 
Here, the left and right leads considered in the WP simulations have separate absorptive layers consisting of 100 sites, with exponential decay rates $\gamma_n = 2(1 - \exp(-n^2/80))\, \mathrm{eV}$ for sites $n\in\{1,2,\ldots,100\}$ within the layer. 
Excluding the absorptive layer, the number of sites in the left lead is set to $10\Delta$ to fully contain the initial Gaussian state, while the right lead is modeled by 200 sites regardless of $\Delta$. 
The initial wave packet is centered at the midpoint of the left lead and is evolved, using exact diagonalization, to the final time $t_\text{max} \approx 33\, \mathrm{ps}$. 
The calculations are performed on a uniform energy grid of 1001 points over the interval $\left[\epsilon_L - 2|t_L|,\ \epsilon_L + 2|t_L|\right]$. 
Note that the WP results converge to the NEGF results when the width of the initial Gaussian state is sufficiently large (i.e., $\Delta \gg 10$). 
These results demonstrate that the WP method can accurately reproduce the NEGF results when the initial Gaussian state has a well-defined energy with sufficiently small energy uncertainty.

\begin{figure*}
    \centering
    \includegraphics[width=\textwidth]{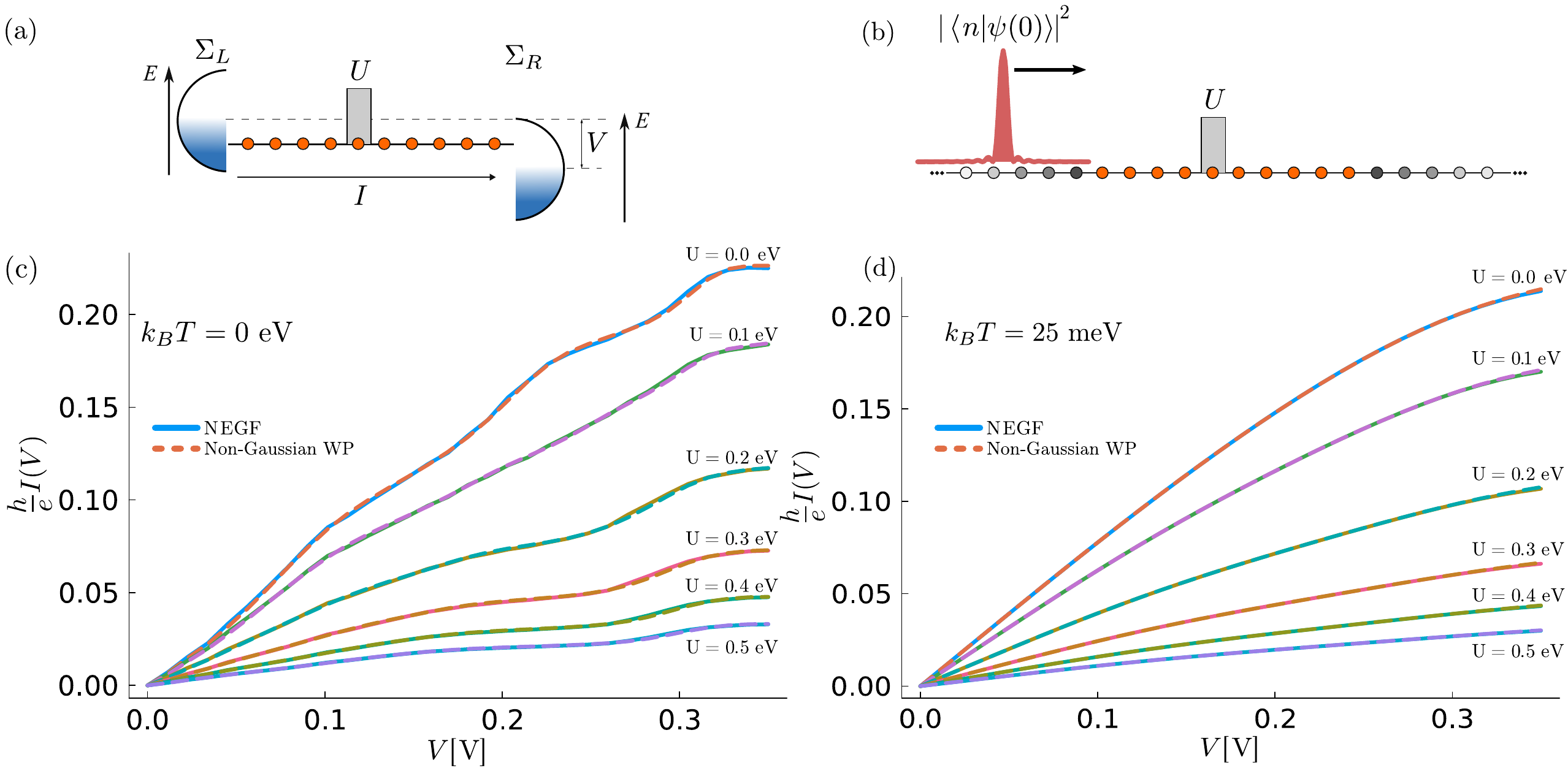}
    \caption{Current-voltage characteristics computed using the NEGF and \emph{non-Gaussian} WP methods, schematically represented in (a) and (b), respectively. The molecular system consists of $N=10$ sites, where an impurity is introduced at the middle site, shifting the on-site energy to $\epsilon_n = \epsilon_0 + U \delta_{n, 5}$. Other simulation parameters are the same as in Fig.~\ref{fig:WP-electron}. 
    The NEGF (solid lines) and non-Gaussian WP results (dashed lines) are well matched for several barrier heights $U \in\{0, 0.1, 0.2, 0.3, 0.4, 0.5\}\,\mathrm{eV}$ at both (c) zero and (d) room temperatures.}
    \label{fig:IV-barrier}
\end{figure*}

Conversely, the WP results obtained using a finite width $\Delta$ of the initial Gaussian state can be reproduced by the NEGF method. 
In Fig.~\ref{fig:WP-electron}(b), we show that the transmission probability $T_{W}(E)$ computed by the WP method with a fixed width $\Delta = 10$ is well matched to the Gaussian-weighted average of the NEGF transmittance $T_{N}(E')$, obtained using Eq.~(\ref{eq:WP_NEGF_Gaussian}).

These results demonstrate that the energy-resolved transmittance computed using NEGF can be reproduced by the WP method with a sufficiently broad initial Gaussian state, and that the WP results obtained with a finite-width initial Gaussian state can, in turn, be reproduced by appropriately averaging the NEGF transmittance.

\subsection{Current-Voltage Characteristics}\label{sec:results_current}

\begin{figure*}
    \centering
    \includegraphics[width=\textwidth]{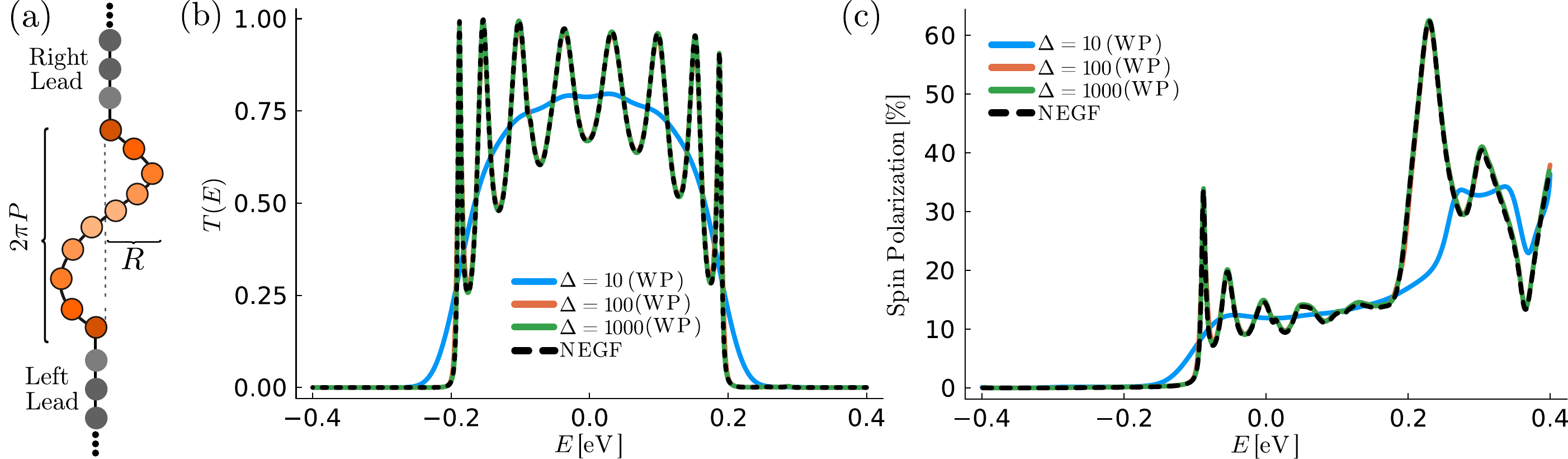}
    \caption{(a) Schematic of a helical molecule with radius $R$ and pitch $2\pi P$ connected to input (left) and output (right) leads. (b) Total transmittance of an electron in the presence of spin-phonon interaction, computed using NEGF (black dashed line) and WP (colored solid lines) methods with controlled width $\Delta\in\{10,100,1000\}$ of the initial Gaussian state. The NEGF and WP results are well matched for $\Delta \gg 10$. (c) Corresponding spin polarizations. In simulations, we consider $\epsilon_L = \epsilon_R = \epsilon_n = 0$, $t_L = t_R = 0.2\, \mathrm{eV}$, $t_n = 0.1\, \mathrm{eV}$, $g_L = g_R = 0.2\, \mathrm{eV}$, $\omega=0.1\,{\rm eV}$, $\alpha=0.01\,{\rm eV}$, $R=0.7\,{\rm nm}$, $2\pi P=3.4\,{\rm nm}$, and zero temperature.}
    \label{fig:vibronic}
\end{figure*}

We now show how an energy-integrated steady-state observable can be obtained as an asymptotic quantity from a time-domain state with a suitably tailored spectrum. Specifically, we compute the current directly within the WP framework using a non-Gaussian initial state whose spectral weight reproduces the Landauer transport window.
We assume that, under an applied bias voltage $V$, the site energies of the input and output leads are shifted by $\epsilon_L=\epsilon_0+eV/2$ and $\epsilon_R=\epsilon_0-eV/2$, such that the eigenvalue spectra of the left and right leads become ${E = \epsilon_L+2t_L\cos(k_L)}$ and ${E = \epsilon_R+2t_R\cos(k_R)}$, respectively. When the eigenstates of both leads are at half filling, as schematically illustrated in Fig.~\ref{fig:IV-barrier}(a), the current $I$ at a given voltage $V>0$ is computed within the NEGF formalism using the Landauer-B{\"u}ttiker formula
\begin{align}
    I = \frac{e}{h}\int_{-\infty}^{\infty}dE \,T_{N}(E)(f_L(E)-f_R(E)),\label{eq:LBformula}
\end{align}
where $f_{j}(E)=({\rm e}^{(E-\epsilon_{j})/k_B T}+1)^{-1}$ denotes the Fermi-Dirac distribution of lead $j\in\{L,R\}$, satisfying ${f_L(E)-f_R(E)\ge 0}$.
For $V < 0$, the sign of ${f_L(E) - f_R(E)}$ is reversed, and so is the direction of the current $I$. 
Leveraging Eq.~(\ref{eq:WP_NEGF_E}), we assume that the WP transmittance can be expressed as an integral of the NEGF transmittance with an energy-dependent weighting factor
\begin{align}
    T_{W} &= \int_{-\infty}^{\infty}dE \,T_{N}(E)(f_L(E)-f_R(E))\label{eq:T_W_nonGaussian}\\
    &= \int_{-\infty}^{\infty} dE\,T_{N}(E)\frac{|B_{k(E)}|^2}{2|t_L \sin(k(E))|},
\end{align}
with $E=\epsilon_L+2t_L\cos(k)$, which is proportional to the current in Eq.~(\ref{eq:LBformula}). 
Note that even if the NEGF transmittance is unity for all energies, i.e., $T_{N}(E)=1$ for all $E$, the integrated value $A=\int_{-\infty}^{\infty}dE (f_L(E)-f_R(E))$ is not unity. 
Moreover, both $A$ and $T_W$ in Eq.~(\ref{eq:T_W_nonGaussian}) have units of energy. 
In this work, we use this value $A$ to normalize the initial non-Gaussian state in the WP method, defined as
\begin{align}
    \ket{\psi(0)}&\propto \int_{0}^{\pi} dk B_k \sum_{n}{\rm e}^{{\rm i}k(n-n_0)}\ket{n},\label{eq:non-gaussian-1}\\
    B_k&=\sqrt{2|t_L\sin(k)(f_L(E)-f_R(E))|}, \label{eq:non-gaussian-2}
\end{align}
where $E=\epsilon_L+2t_L\cos(k)$, $\langle\psi(0)|\psi(0)\rangle=A$, and $n_0$ is the center position of the initial state where the input plane waves ${\rm e}^{{\rm i}k(n-n_0)}$ interfere constructively.
Such a non-Gaussian initial WP is represented in Fig.~\ref{fig:IV-barrier}(b).

We note that the non-Gaussian state in \eqref{eq:non-gaussian-1} should be viewed not as a physical, unnormalized, single-electron wavefunction but rather as an auxiliary state for evaluating the Landauer integral.
Since the Hamiltonian is time independent, different energy components scatter independently, and the asymptotic flux is diagonal in energy basis. 
Consequently, up to the prefactor $e/h$, the transmitted norm from this auxiliary state evaluates the same energy integral as the Landauer-Büttiker formula in \eqref{eq:LBformula}.

In Figs.~\ref{fig:IV-barrier}(c) and (d), we show the current-voltage characteristics at zero and room temperatures, respectively, computed using the NEGF and non-Gaussian WP methods for a molecular system consisting of $N=10$ sites, where an impurity is introduced at the middle site, shifting the on-site energy to $\epsilon_n = U\delta_{n,5}$. In the WP simulations, we consider a left lead consisting of 200 sites, where the initial non-Gaussian state is centered at site $n_0 = 70$, and a right lead consisting of 100 sites. Both leads have a 70-site absorptive layer with quadratic decay rates $\gamma_n = \gamma n^2$ with $\gamma = 10^{-5}\, \mathrm{eV}$ for sites $n\in\{1,2,\ldots,70\}$ within the layer.
The time evolution is computed until $t_\text{max} = 527\, \mathrm{fs}$ using the RK4 method with a timestep $\Delta t = 0.07\, \mathrm{fs}$.
The NEGF calculation are performed on a uniform energy-grid of 1001 points in the interval $\left[\epsilon_0 - (2|t_L| + eV/2),\  \epsilon_0 + (2|t_L| + eV/2)\right]$.
The quantitative agreement between the two methods across ranges of bias voltage $V$ and barrier heights $U$ both at zero and room temperatures demonstrates that the WP method can directly compute the current for a given voltage by preparing a non-Gaussian initial state, without evaluating energy-resolved transmittance, in contrast to the NEGF method.

\subsection{Spin Polarization}\label{sec:results_SP}

So far we have considered a simple one-dimensional electron model without spin and other physical degrees of freedom.
We now show that this correspondence also holds when the molecular region includes additional internal degrees of freedom, inelastic scattering channels, and when the observable of interest is spin-dependent.
We consider the spin-phonon model described by Eq.~(\ref{eq:H_S_spin_phonon}), in which spin-dependent interaction occurs between the electron and the phonon mode within the molecular region, while the leads have no spin-phonon coupling.
As schematically shown in Fig.~\ref{fig:vibronic}(a), the radius and pitch of the helical molecule consisting of $N = 10$ sites are taken as $R=0.7\ \mathrm{nm}$ and $2\pi P = 3.4\ \mathrm{nm}$, respectively, which are typical values for DNA. 
The phonon mode is modeled with $\omega = 0.1\ \mathrm{eV}$ and $\beta = 1/(2D)$, where $D = \sqrt{R^2+P^2}$, and the spin-phonon coupling is set to $\alpha = 0.01\ \mathrm{eV}$. 
For simplicity, we consider zero temperature, so the phonon mode is initially in its vacuum state. 
The site energies and inter-site couplings are the same as in Fig.~\ref{fig:WP-electron}. 
In Fig.~\ref{fig:vibronic}(b), the total transmittance $T_\uparrow + T_\downarrow$ is shown as a function of electron energy $E$, where up to 50 phonon levels (i.e., $m \in \{0,1,\ldots,49\}$) are considered to ensure convergence of the simulated results. 
The other simulation parameters are the same as in Sec.~\ref{sec:results_trans}.
For a sufficiently large width $\Delta$ of the initial Gaussian state in the WP method, the WP results agree well with the NEGF results obtained using the formalism for the multi-terminal setup summarized in Sec.~\ref{sec:NEGF_MT}. 
The spin polarizations ${\mathrm{SP}=(T_\uparrow-T_\downarrow)/(T_\uparrow+T_\downarrow)}$ obtained by the two methods are also well-matched, as shown in Fig.~\ref{fig:vibronic}(c).

\section{Conclusions}\label{sec:conclusions}

In this work, we have established an explicit correspondence between stationary energy-resolved transport and finite-size wave packet propagation, in which the spectral content of the incoming WP determines the energy weighting of the corresponding asymptotic transmission observable.
We have first used this correspondence to demonstrate that the electron transmittance computed using the steady-state NEGF method can be quantitatively reproduced by the WP approach, provided that the initial Gaussian wave packet has a well-defined energy with sufficiently small energy uncertainty.
Conversely, we have shown that the transmittance obtained from the WP method can be recovered by appropriately averaging the NEGF transmittance over energy. 
Importantly, we then used this correspondence to construct a WP encoding the Landauer energy-window.
The propagation of this non-Gaussian initial WP directly computes the current at a given voltage, without evaluating energy-resolved transmittance, in contrast to the NEGF method.
Further spin-phonon calculations show that the correspondence extends to spin-resolved transport in the presence of inelastic scattering channels relevant to CISS.
These results establish the consistency of steady-state NEGF and WP methods, and the corresponding asymptotic observables, for common microscopic models.
In the context of CISS, this work helps to separate the effects of the transport formalism from those of the underlying microscopic model and physical assumptions.

Our findings highlight the complementary nature of the two frameworks for studying electron transport in molecular systems. 
The NEGF method provides high-resolution, energy-resolved steady-state transmittance, whereas the WP approach offers direct access to transient dynamics, enabling detailed insights into the internal evolution of electronic states.
In this context, combining the WP framework with model parameters derived from \emph{ab initio} methods, such as density functional theory, represents a promising direction for elucidating the microscopic mechanisms underlying enantiospecific electron transport in chiral molecular systems.

\section*{Acknowledgements}

This work was supported by the ERC Synergy grant HyperQ (Grant No.~856432), the BMBF via project PhoQuant (Grant No.~13N16110), the Volkswagen Foundation (Grant No.~0200187), the state of Baden-W{\" u}rttemberg through bwHPC and the German Research Foundation (DFG) through Grant No.~INST 40/575-1 FUGG (JUSTUS 2 cluster).

\end{document}